\documentclass[AMA,STIX1COL]{WileyNJD-v2}

\articletype{Article Type}%

\received{26 April 2016}
\revised{6 June 2016}
\accepted{6 June 2016}

\usepackage{placeins}
\usepackage[super]{nth}
\usepackage{subcaption}

\begin{document}

\title{Spectral/hp element simulation of flow past a Formula One front wing: validation against experiments}

\author[1,2,3]{Filipe F. Buscariolo}
\author[2]{Julien Hoessler}
\author[4]{David Moxey}
\author[5]{Ayad Jassim}
\author[1]{Kevin Gouder}
\author[1]{Jeremy Basley}
\author[1]{Yushi Murai}
\author[3]{Gustavo R. S. Assi}
\author[1]{Spencer J. Sherwin*}

\authormark{Filipe F. Buscariolo \textsc{et al}}

\address[1]{\orgdiv{Department of Aeronautics}, \orgname{Imperial College London}, \orgaddress{\state{London}, \country{United Kingdom}}}

\address[2]{\orgdiv{CFD Methodology Group}, \orgname{McLaren Racing}, \orgaddress{\state{Woking}, \country{United Kingdom}}}

\address[3]{\orgdiv{N\'{u}cleo de Din\^{a}mica e Fluidos Research Group}, \orgname{Univeristy of S\~{a}o Paulo}, \orgaddress{\state{S\~{a}o Paulo}, \country{Brazil}}}

\address[4]{\orgdiv{College of Engineering, Mathematics and Physical Sciences}, \orgname{University of Exeter}, \orgaddress{\state{Exeter}, \country{United Kingdom}}}

\address[5]{\orgdiv{SGI}, \orgname{Hewlett Packard Enterprise}, \orgaddress{\state{Reading}, \country{United Kingdom}}}

\corres{*Spencer J. Sherwin, Department of Aeronautics, Imperial College London, London, SW7 2AZ, United Kingdom \email{s.sherwin@imperial.ac.uk}}

\presentaddress{Department of Aeronautics, Imperial College London, London, SW7 2AZ, United Kingdom}

\abstract[Summary]{Emerging commercial and academic tools are regularly being applied to the design of road and race cars, but there currently are no well-established benchmark cases to study the aerodynamics of race car wings in ground effect. In this paper we propose a new test case, with a relatively complex geometry, supported by the availability of CAD model and experimental results. We refer to the test case as the Imperial Front Wing, originally based on the front wing and endplate design of the McLaren 17D race car. A comparison of different resolutions of a high fidelity spectral/hp element simulation using under-resolved DNS/implicit LES approach with fourth and fifth polynomial order is presented. The results demonstrate good correlation to both the wall-bounded streaklines obtained by oil flow visualization and experimental PIV results, correctly predicting key characteristics of the time-averaged flow structures, namely intensity, contours and locations. This study highlights the resolution requirements in capturing salient flow features arising from this type of challenging geometry, providing an interesting test case for both traditional and emerging high-fidelity simulations.}



\maketitle


\section{Introduction}
\label{sec:Introduction}

External aerodynamics is one of the key performance differentiators in Formula One: a mere 1-2\% improvement in aerodynamics performance can be the difference between first and tenth.  As a result, aerodynamics has become one of the main design focal points for Formula One performance and clearly involves challenging fluid mechanics. 

\begin{figure}[bt]
	\begin{center}
	\includegraphics[width=0.5\textwidth]{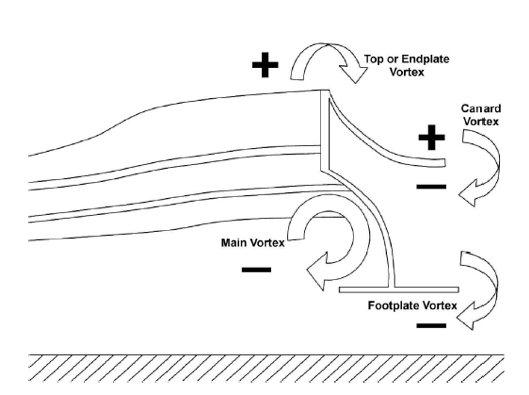}
	\caption{Topology of the complex vortex system downstream of the front wing. Reproduced from Pegrum~\cite{pegrum2007experimental}.}
	\label{fig: 1}
	\end{center}
\end{figure}

For open wheel racing cars, the front wing is a key aerodynamic feature whose role is twofold: the generation of load on the front axle (i.e. downforce), and secondly, as presented on Figure \ref{fig: 1}, the production of a vortical system to control the flow downstream, and in particular to limit the negative effect of the open front wheel wake on the rest of the car.

Although there are a number of Formula One test cases in the open literature \cite{zerihan2000aerodynamics,zhang2003aerodynamics,ahmed2007aerodynamics}, these have typically been constructed to be visually similar to existing race cars but have no experimental validation. In this paper we propose a test case, referred to as the Imperial Front Wing (IFW) model, which is based on the McLaren 17D race car front end. The geometry was previously investigated experimentally by Pegrum~\cite{pegrum2007experimental}. As shown in the sketch in figure \ref{fig: 1}, the geometry consists of a multi-element wing in ground effect, leading to the generation of a series of interacting vortical structures, thereby creating a highly complex flow environment downstream of the wing. Accurately modelling these vortices and their interactions, by capturing their relative position and strength, is of utmost importance for a successful front wing design. 

Having an efficient and reliable computational fluid dynamics (CFD) toolchain is critical in reducing time and cost for the development of road and racing cars. The current methods available, in ascending order of computational cost and accuracy, are Reynolds Averaged Navier-Stokes (RANS), unsteady RANS, detached eddy simulation (DES), wall modelled large eddy
simulation (WMLES), implicit large eddy simulation -- sometimes referred to as under-resolved direct numerical simulation (iLES/uDNS) -- and direct numerical simulation (DNS).

With the advancements of computing capability, an emerging trend in CFD methodology is the application of high-order methods, such as spectral/hp element methods. This discretisation takes advantage of the high accuracy and advantageous convergence properties of spectral (p) methods, while retaining the flexibility of the classical finite element (h) method, allowing complex geometries to be efficiently captured. After decades of developments, significant progress has been made to the technological readiness of these methods. CFD software suites, such as Nektar++\cite{cantwell2015nektar++}, have already made high-order spectral/hp element methods widely accessible to a broad community of users.

In this paper, we compare the simulation results obtained from Nektar++ using various polynomial orders of the spectral/hp element method, with experimental measurements of surface flow visualisation and time resolved particle image velocimetry (PIV). Solutions of high accuracy were directly obtainable thanks to the implementation of several novel techniques for the stabilisation of higher-order spectral/hp methods. RANS simulation were also conducted to initialise the spectral/hp simulation.

The paper is organised as follows.
The geometry of the IFW test case is presented in Section~\ref{sec:IFW}, followed by the description of the experimental setup in Section~\ref{sec:ExperimentalSetup}. In Section~\ref{sec:hpSimulation}, we outline the spectral/hp numerical simulation configurations. The comparisons between different resolution simulations and the experimental measurement are discussed in Section~\ref{sec:Results}, followed by concluding remarks in Section~\ref{sec:Conclusions}.

\FloatBarrier

\section{Imperial Front Wing}
\label{sec:IFW}

\begin{figure}[bt]
	\begin{center}
	\includegraphics[width=0.50\textwidth]{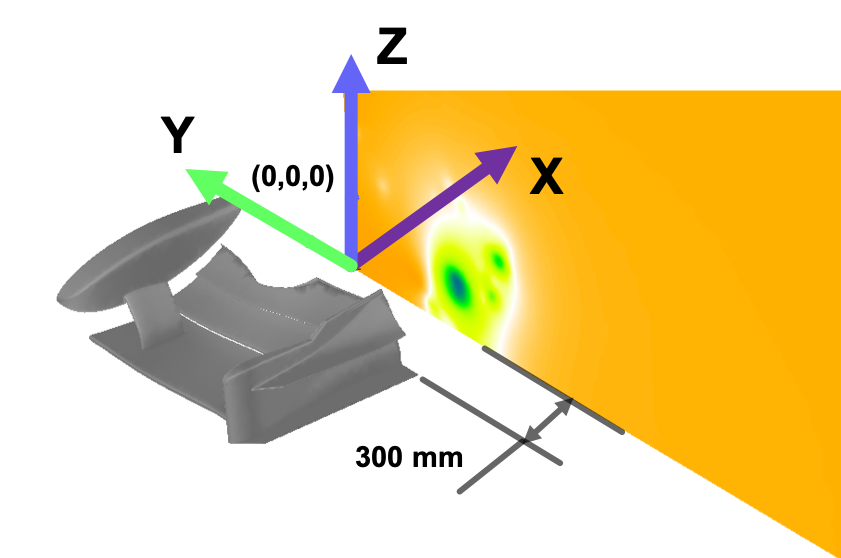}
	\caption{Imperial Front Wing geometry with its coordinate axis definition.}
	\label{fig: 2}
	\end{center}
\end{figure}

As previously mentioned, the IFW test case is based on the McLaren 17D Formula One race car and its geometry consists of a three-element front wing with an endplate attached to a simplified nose cone. In the following, we will use a coordinate system with $X$ denoting the streamwise direction, $Y$ the spanwise direction and $Z$ the vertical direction. The trailing edge of the front wing is located at -300mm in $X$ direction and the ground plane at -25mm in the $Z$ direction. A 3D illustration is presented in Figure~\ref{fig: 2} and the geometry CAD files are available at DOI: 10.14469/hpc/6049.
The wing is two-sided, and PIV measurements were only performed on the $y<0$ endplate. 

Despite looking simpler than present-day front wings, which have benefited from another fifteen years of development, the IFW captures the essential aerodynamic features of a front wing configuration, and still poses significant aerodynamics challenges in accurately estimating the aerodynamic loads, the locations of transition lines, the generation and shedding pattern of multiple vortices and their respective strengths.

\section{Experimental setup}
\label{sec:ExperimentalSetup}

\subsection{Wind tunnel details}
\label{subsec:DomainDetails}

\begin{figure}[bt]
	\begin{center}
	\includegraphics[width=0.45\textwidth]{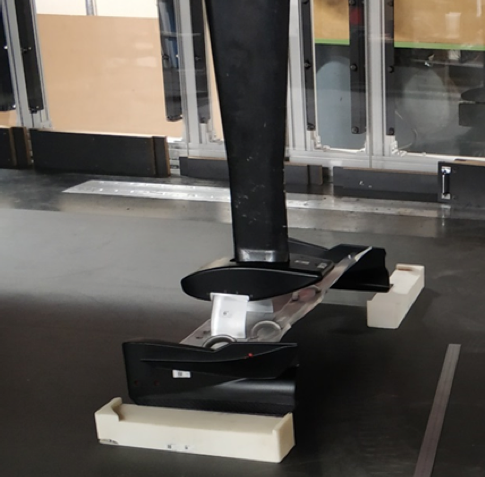}
	\caption{Imperial Front Wing assembly on Imperial College London 10~$\times$~5 wind tunnel.}
	\label{fig:windtunnel}
	\end{center}
\end{figure}

Wind tunnel experiments were performed in the 10~$\times$~5 Wind Tunnel in the Department of Aeronautics at Imperial College London \cite{imperial10times5}. It has a closed return circuit with closed twin test sections. The lower test section is 20m long with a cross sectional area of 3m~$\times$~1.5m, equipped with a rolling road and full boundary layer control. The test section used for this experiment is temperature controlled with flow uniformity around 1.0\% and turbulence intensity less than 0.25\%. The experimental model is of 1:2 scale, assembled in the test section as shown on Figure \ref{fig:windtunnel}.

\subsection{Domain parametrisation}
\label{subsec:DomainParametrisation}

\begin{figure}[bt]
	\begin{center}
	\includegraphics[width=0.5\textwidth]{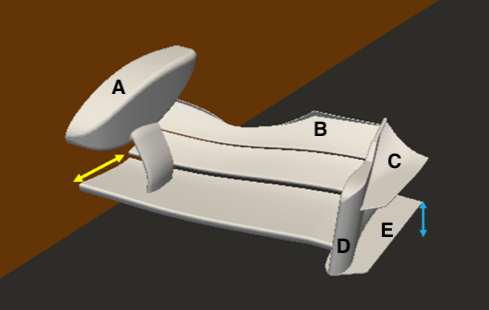}
	\caption{Imperial Front Wing geometry. 
Nomenclature: ride height $h$ (light blue line), chord $c$ (yellow line), main plane (dark orange), moving belt (dark grey), nose cone (A), Gurney flap (B), canard (C), endplate (D) and footplate (E).}
	\label{fig:notation}
	\end{center}
\end{figure}

As shown in figure \ref{fig:notation}, we denote by $h$ the ride-height (i.e.\ the distance between the ground and the lowest part of the front wing endplate) and by $c$ the chord length of the main element. The position of the wing in the tunnel is further characterised by a pitch angle of 1.094$^{\circ}$. 

This study uses the configuration of $h/c=0.36$ which can be considered as a relatively low front ride height, with high ground effect and hence higher loads on the wing. The corresponding Reynolds number is $Re = 2.2 \times 10^{5}$, based on the main element chord $c$ of 250mm and a free stream velocity $U$ of 25m/s.

\subsection{Measurements}
\label{subsec:Measurements}

The experiments performed in this study consist of flow visualizations of wall-bounded streaklines using oil flow measurements, and time resolved PIV on distinct planes of various locations, as presented below.

\subsubsection{Surface flow patterns}
\label{subsubsec:SurfaceFlowPatterns}

The first quantity of interest is the analysis of wall-bounded streaklines on the wing. To that end, flow visualization paint was applied on the wing in order to identify separation and transitional regions, at a reduced velocity of 15m/s (i.e. $Re = 1.32 \times 10^{5}$). Visualization results will be compared to wall shear stress contours from the numerical simulations at $Re=2.2 \times 10^{5}$. To make the experiment easier to reproduce numerically, unlike previous experiments by \cite{pegrum2007experimental}, we elected not to use any transition strips, hence adopting a natural transition mechanism.

\subsubsection{Particle image velocimetry}
\label{subsubsec:PIVMeasurements}

\begin{figure}[bt]
	\begin{center}
	\includegraphics[width=0.4\textwidth]{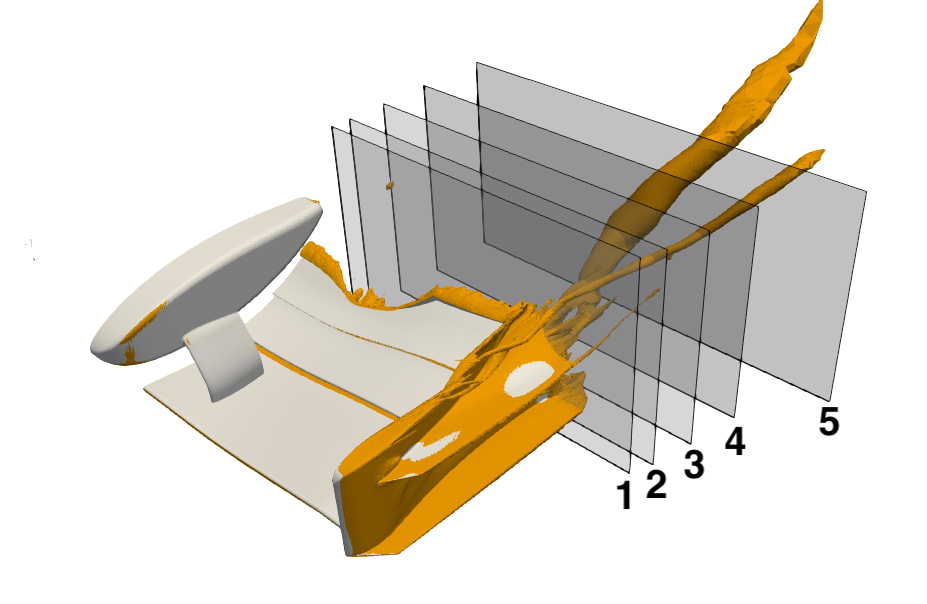}
	\caption{Imperial Front Wing geometry and selected PIV planes.}
	\label{fig: planes}
	\end{center}
\end{figure}

To capture the evolution of vortices travelling downstream, Particle Image Velocimetry (PIV) measurements were performed in various iso-$X$ planes within an inspection window of 450mm~$\times$~260mm. The inspection plane closest to the trailing edge of the wing (located at $X$=-300mm)  was at $X$=-294mm. It was not possible to get any closer to the trailing edge without being affected by laser reflections. It should however provide a good representation of the initial position of the vortical structures leaving the wing. A further four inspection planes were located respectively at $X=-250mm,-150mm,-24mm$ and $150mm$. For each plane, the total averaging time was 10 seconds, at an acquisition frequency of 250Hz. The locations of the interrogation planes (also normalised by the main element chord) are indicated in Table~\ref{tab:PlaneLocations} and Figure~\ref{fig: planes} and are based on the coordinate system in Figure~\ref{fig: 2}.



\begin{table}[b]
\begin{center}
\begin{tabular}{crr}
\toprule
\textbf{Plane Number} & \textbf{$X$ [mm]} & \textbf{$X/c$ [-]}\\
\midrule
1 & -294 & -1.176\\
2 & -250 & -1.000\\
3 & -150 & -0.600\\
4 & -24 & -0.096\\
5 & 150 & 0.600\\
\bottomrule
\end{tabular}
\end{center}
\caption{Plane locations for PIV measurements}
\label{tab:PlaneLocations}
\label{time}
\end{table}

\section{Spectral/hp uDNS/iLES Simulation}
\label{sec:hpSimulation}

High fidelity numerical simulation were performed using the implicit LES formulation based on a spectral/hp element discretisation provided in the Nektar++ open source software~\cite{cantwell2015nektar++}. In the spectral/hp method, the domain is first divided into non-overlapping elements, similar to a finite volume or finite element methods, offering geometric flexibility and allowing for local refinement. The solution in each element is then approximated by high order polynomial expansion.

All simulations were performed using the incompressible Navier-Stokes solver which employs a velocity correction scheme \cite{guermond2003velocity}. The elliptic operators were discretised using a classical continuous Galerkin (CG) formulation whereas the advection operator on the formulation used a discontinuous Galerkin (DG) projection to allow for a sub-stepping algorithm as proposed by \cite{sherwin2003substepping}. For well-resolved simulations, it is often possible to discretise the pressure and velocity space using the same polynomial order; however this problem formulation does not satisfy the Ladyzhenskaya-Babu\v{s}ka-Brezzi (LBB) or inf-sup condition, thus numerical stability and convergence are not guaranteed~\cite{ferrer-2014}. 

It is more appropriate to specify an equivalent of the Taylor-Hood approximation and use one polynomial order higher for the velocity fields than the pressure field with a continuous expansion. Therefore in what follows, our results are named using two digits to quantify both the polynomial order of velocity and pressure. For example, a simulation referred to as NM43 will have a polynomial order of 4 for velocity and 3 for pressure. In addition, when solely referring to the simulation expansion order, we will typically mean the order related to the velocity variable. 

At higher Reynolds numbers, where the flow is typically only marginally resolved, we regularly observe numerical instabilities related to wave interaction and wave trapping. To mitigate these effects, we employ both dealiasing and Spectral Vanishing Viscosity (SVV) stabilization techniques. For the SVV operator in this study, we run the simulation using a novel CG-SVV scheme with a DG Kernel as proposed in~\cite{moura2017eddy}. This is the first time results using this formulation have been verified and validated against experimental data for 3D simulations at high Reynolds number. The fundamental idea is based on fixing the P\'eclet number, which can be understood as a numerical Reynolds number based on local velocity and mesh spacing for the whole domain. This is achieved by making the viscosity coefficient of the SVV operator proportional to both a representative velocity and a local measure of mesh spacing. Once the P\'eclet number is the same for the domain, \cite{moura2017eddy} proposed a SVV kernel operator for CG methods that mimics the properties of discontinous Galerkin (DG) discretisations, which exhibits natural damping of high frequencies and reflected waves. In this approach the dissipation curves arising from spatial eigenanalysis of CG of order $p$ are matched to those of DG with order $p-2$. Matching both curves offers benefits for simulations at very high P\'eclet number.

Finally, aliasing error related to the Navier-Stokes equations appears when handling its quadratic non-linearity term by using the Gauss integration orders $Q$ similar to the solution polynomial order $P$. This is usually present in simulations considering under-resolved turbulence, such as iLES, which leads to a significant error increment in high-frequency modes of the solution and can cause the simulations to diverge. To avoid these errors, we employed a quadrature order consistent with the polynomial order and quadratic non-linearity of the Navier-Stokes equations. In areas of non-linear geometry deformation, we also have to be mindful of geometric aliasing which arises from the geometric mapping used to deform elements to align with the underlying CAD surfaces. We refer the interested reader to \cite{mengaldo2015dealiasing} for more details. 

\subsection{Boundary conditions}
\label{subsec:BoundaryConditions}

The boundary conditions for the computational study were as follows:
\begin{itemize}
\item Both front wing and nose cone are set as wall with no-slip condition;
\item A half model of the geometry is used with a symmetry condition imposed at $Y = 0$;
\item Uniform velocity profile is imposed at the inlet;
\item A high-order outflow condition is imposed at the outlet, as proposed by \cite{dong2014robust};
\item A moving ground condition is imposed on the floor with speed $U$ in the $X$ direction.
\end{itemize}

\subsection{Mesh generation}
\label{subsec:MeshGeneration}

\begin{figure}[bt]
	\begin{center}
	\includegraphics[width=0.6\textwidth]{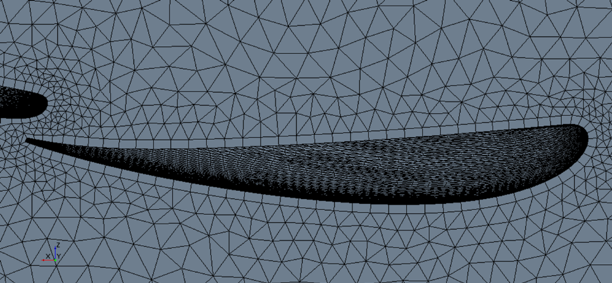}
	\caption{Detailed view of the macro prism layer.}
	\label{fig: 4}
	\end{center}
\end{figure}

The mesh for this study in Nektar++ is first generated with a finite volume (FV) mesher using a conformal hybrid tetahedron/prism layout with a single `macro' prism layer of 3mm. The initial mesh is shown in Figure~\ref{fig: 4}, where we observe the macro mesh around the main wing element on the symmetry plane. We note that this mesh is much coarser when compared with a standard finite volume models with similar solution accuracy, recalling that within each element a polynomial expansion is applied in the spectral/hp element methods.

\begin{figure}[bt]
	\begin{center}
	\includegraphics[width=0.6\textwidth]{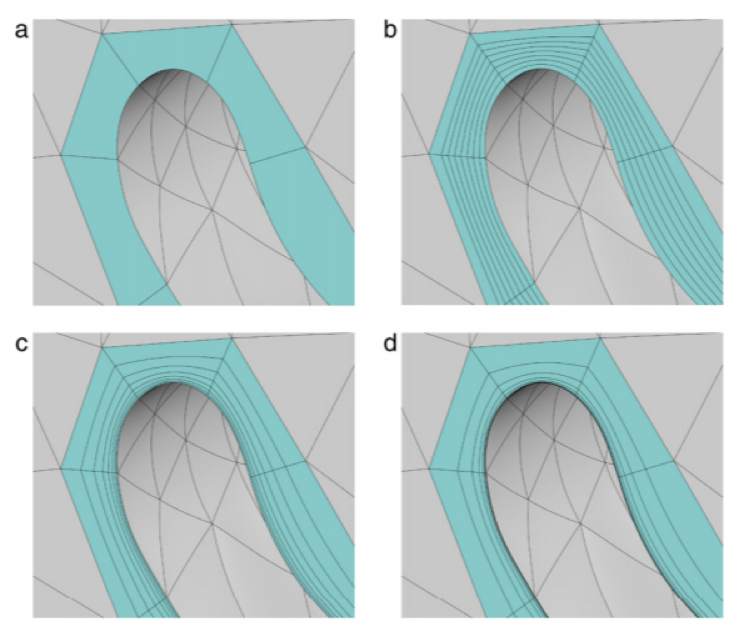}
	\caption{A sequence of meshes obtained by splitting macro-elements into $n$ = 8 elements using a geometric progression and various values of growth rate $r$. (a) The macro-element mesh; (b) $r = 1$; (c) $r = 3/2$; (d) $r = 2$. Reproduction courtesy of Moxey \cite{moxey2015isoparametric}}
	\label{fig: 5}
	\end{center}
\end{figure}

In order to make the mesh suitable for higher orders, as well as to provide a mesh with an appropriate boundary layer spacing as required at higher Reynolds numbers, the mesh is then processed through NekMesh~\cite{turner2017high}, a high-order mesh generator/modifier for Nektar++. We note that where a high-order element meets a CAD surface, it must be deformed in order to align with the CAD and thus faithfully represent the geometry. The use of NekMesh applies surface curvature in a manner so as to minimise the projection error as described in~\cite{turner2017high}, and the prism layer is split using an isoparametric splitting technique developed in \cite{moxey2015isoparametric}. The outcomes of this process with different growth rate settings for the boundary layer progression are illustrated in figure~\ref{fig: 5}.

For the splitting of the macro prism layer on both the front wing and the moving ground, we subdivide the prism layer into 7 elements ($n=7$) using a geometrical growth rate ($r$) of 1.6. For all simulation cases, the high-order mesh implemented is of \nth{6} order. These parameter settings for the mesh, together with surface and volumetric refinement criteria, are based on the optimization result of an internal study. The final mesh is provided under DOI: 10.14469/hpc/6049 in the Nektar++ XML format. Details of this format can be found under www.nektar.info. 

\subsection{Load case configuration}
\label{subsec:LoadCaseConfiguration}

All simulation are conducted at a nominal velocity $U$ of 25m/s. We have used a mixed expansion order in velocity and pressure where following Taylor-Hood elements, where we recall the pressure expansion is one order lower than the velocity expansion. Initial runs were performed at NM32 (third order expansion in velocity and second order expansion in pressure). We then increased the resolution to NM43, which was the lowest order to obtain reliable results for the mesh discretisation provided with this paper and is discussed in the following sections. To complement these results, we also performed simulations of the IFW with NM54 to evaluate the extend of resolution enhancement with an increment of the polynomial order. 

To save computational time, initialisation techniques are systematically used at various stages of the simulation. We first interpolate the results of a previously converged RANS simulation onto the high order mesh at an expansion order of (NM32), and run this case for two main element chord time units ($t_c$). The unit $t_c$ refers to the number of times flow passes over the chord length $c$ of the main wing element and is defined as $t_c = t_U/c$, where $t_U$ is the total simulation time. These results were then used to initialise the NM43 simulation. This solution was subsequently used as the initial point for higher order simulations. 

\section{Results}
\label{sec:Results}

Before collecting the averages, we first time marched the NM43 for 8 $t_c$, starting from the NM32 restart. This simulation was then time marched a further 2 $t_c$ at NM43, reaching a total of 10 $t_c$. Time averaging of drag and lift coefficients and downstream comparative planes occurs from 8.4 to the 10 $t_c$ for this case. For the \nth{5} order case (NM54) resolution, we restarted the simulation from the NM43 case at 8.4 main element time unit and ran the simulation up to the 11.75 $t_c$ as we verified that we needed to extend further once lift was still evolving with time. The time averaging for the lift and drag at NM54 was started from different times when the signal has reached a stationary state. Therefore, the drag was averaged between the 10.5 and 11.75 $t_c$ and the lift was averaged from 11.5 to 11.75 $t_c$. The comparative planes were averaged from the 8.4 to the 10 $t_c$, similar to the NM43.

\subsection{Wing loading resolution sensitivity}
\label{subsec:LoadConvergence}

\begin{figure}[bt]
	\begin{center}
	\includegraphics[width=0.7\textwidth]{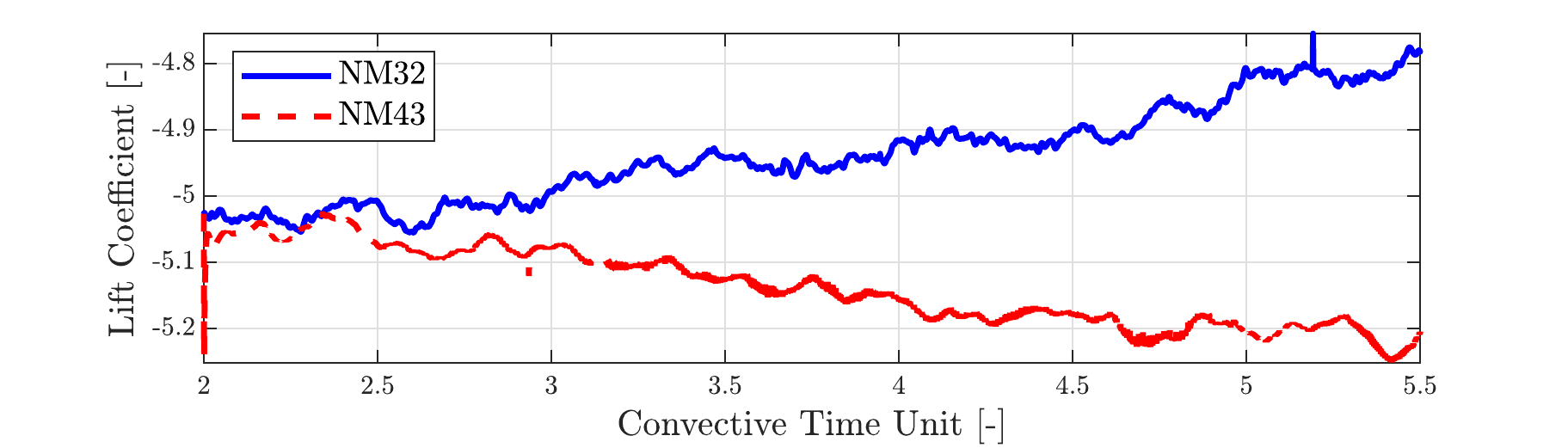}
	\caption{Comparative lift trend line for \nth{3} (NM32 - blue line) and \nth{4} (NM43 - red-dashed line) polynomial expansions obtained from the same NM32 restart file from the \nth{2} to the 5.5 $t_c$} 
	\label{fig:liftNM32}
	\end{center}
\end{figure}

An easily accessible diagnostic when performing the simulations is to consider the lift and drag coefficients. We start by comparing lift/downforce coefficient, based on the main plane surface area, for NM43 and NM32 as shown in Figure~\ref{fig:liftNM32}. We observe there is a very distinct deviation on the lift coefficient where results for NM43 are generally "powering up" (producing more negative lift) whilst the NM32 can be seen to deviate significantly with a reduction in downforce (negative lift). This point will be further discussed in section \ref{subsec:FlowVisualizationResults} when we consider the wall shear stress, however it is important to note that NM32 is an insufficient resolution for this mesh. 

\begin{figure}[bt]
	\begin{center}
	\includegraphics[width=0.7\textwidth]{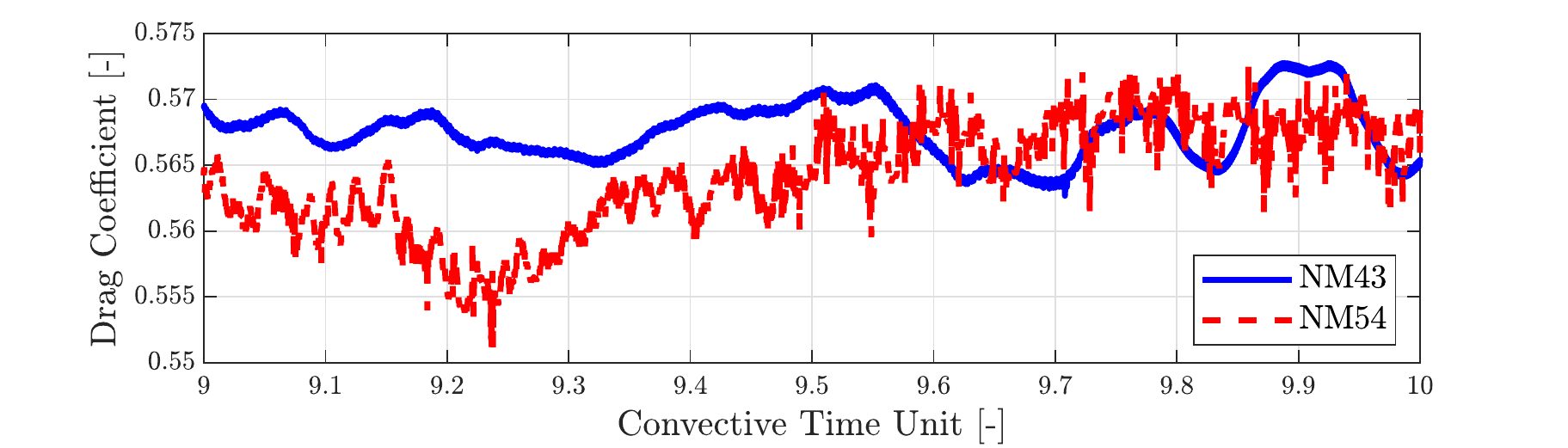}
	\includegraphics[width=0.7\textwidth]{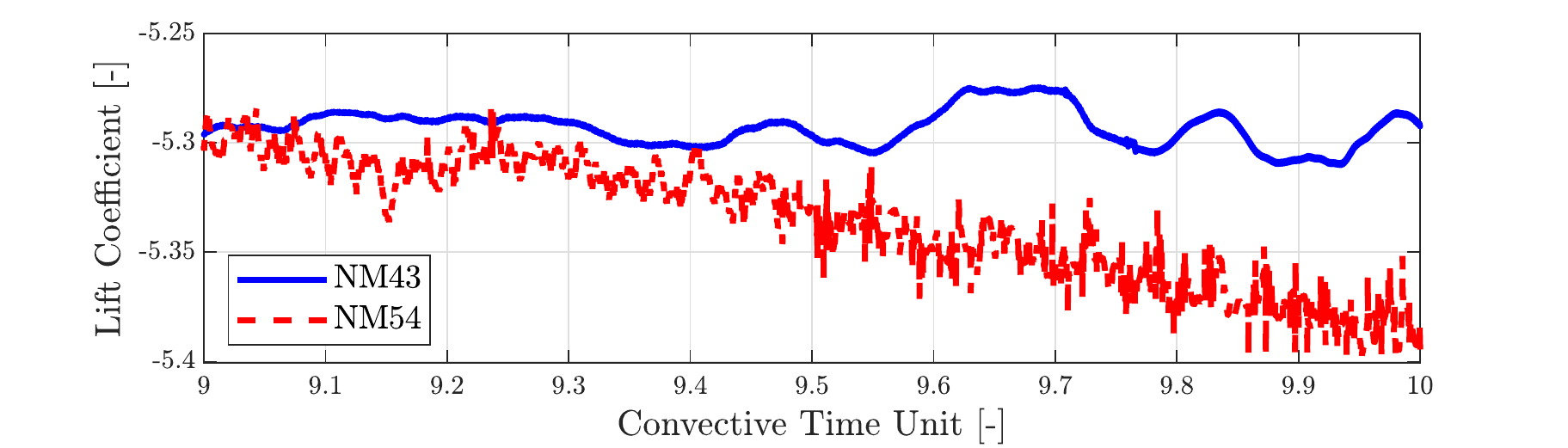}
	\caption{ Drag (top) and Lift (bottom) coefficients history on the IFW comparing NM43 and NM54 results.}
	\label{fig:forces}
	\end{center}
\end{figure}

To demonstrate the convergence of the IFW aerodynamic load with both time and polynomial order, the time history of the drag and lift forces are shown in Figure~\ref{fig:forces}. From this figure we see the evolution of both drag and lift coefficients for the time period from the 8.4 to the 10 $t_c$, based on the main chord for NM43 simulation. We note that both drag and lift coefficients calculation are based on the surface area based on the main wing element chord $c = 0.25$ and the total span of the wing $W_s = 0.7$.

\begin{figure}[bt]
	\begin{center}
	\includegraphics[width=0.7\textwidth]{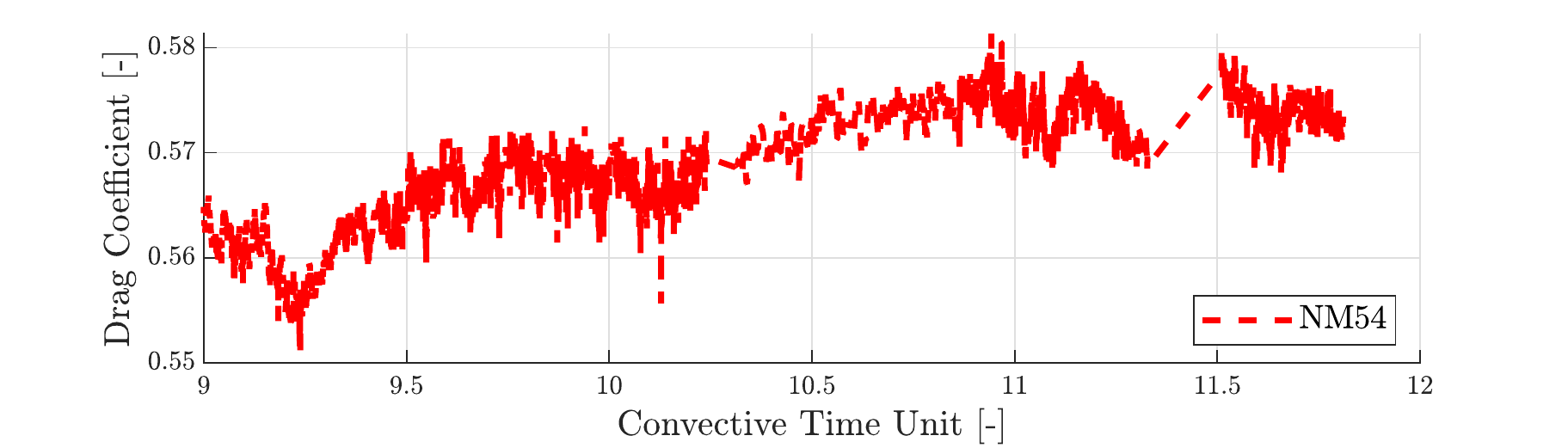}
	\includegraphics[width=0.7\textwidth]{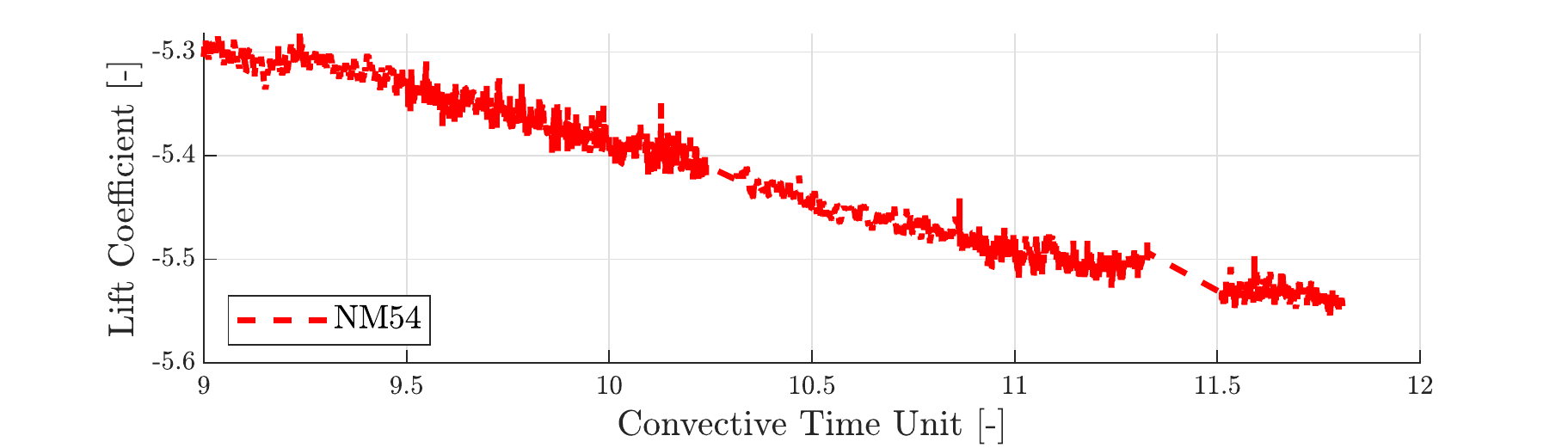}
	\caption{ Drag (top) and Lift (bottom) coefficients history on the IFW for the NM54 results.}
	\label{fig:nm54forces}
	\end{center}
\end{figure}

As seen from Figure~\ref{fig:forces}, the drag and lift coefficients for NM43 simulation have converged to a relatively stationary limit cycle before the 8.4 $t_c$, having well-established behaviour throughout the time window.  In contrast, the drag force for NM54 starts with an oscillatory path and gradually converging to a steady value, higher than the NM43, between the 10.5 and 11.75 $t_c$, as presented on Figure~\ref{fig:nm54forces}. Lift is still evolving all the way to the end of the averaging window. This indicates that the vortices may not have reached their maximum strength in the simulations period however we can see that there is a region between the 11.5 and 11.75 $t_c$ that seems to have a stabilized lift behaviour. Unfortunately limited computational resources prevented us from simulating further.  

Using the above windows for averaging we observe a drag coefficient of {\bf 0.568} at NM43 and {\bf 0.575} at NM54 and a lift coefficient of {\bf -5.30} at NM43 and {\bf -5.54} at NM54. The gap between NM43 and NM54 for drag results is around 1.3\% and 4.5\% for the lift coefficient.

\FloatBarrier

\subsection{Flow Visualization}
\label{subsec:FlowVisualizationResults}

\begin{figure}[bt]
\null\hfill
\begin{subfigure}[b]{0.28\textwidth}
\includegraphics[width=1.0\textwidth]{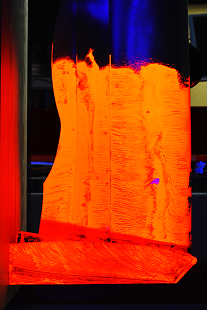}
\end{subfigure}
\hfill
\begin{subfigure}[b]{0.28\textwidth}
\includegraphics[width=1.0\textwidth]{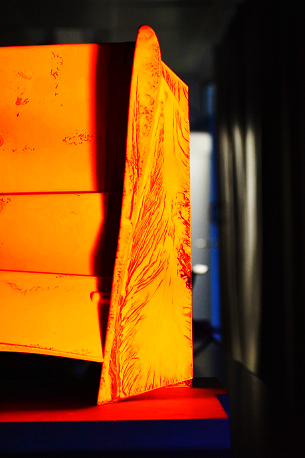}
\end{subfigure}
\hfill
\begin{subfigure}[b]{0.28\textwidth}
\includegraphics[width=1.0\textwidth]{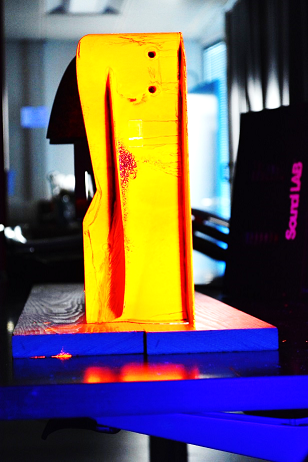}
\end{subfigure}
\hfill\null
\caption{Flow visualization of IFW at velocity of 15m/s using Formula One standard aerodynamic paint, with bottom view (left), top view (middle) and side view (right)}
\label{fig: 7}
\end{figure}

Experimental flow visualization results for IFW are shown in Figure~\ref{fig: 7} in bottom, top and side views respectively. For this experiment, we applied a shear-sensitive type of flow visualization aerodynamic paint on the right-hand side surface of the IFW and performed the test at a lower free stream velocity of 15m/s. The reduced velocity was to avoid distortions in the visualization due to excessive loss of paint, as well as contamination of the wind tunnel.

\begin{figure}[bt]
\null\hfill
\begin{subfigure}[b]{0.28\textwidth}
\includegraphics[width=1.0\textwidth]{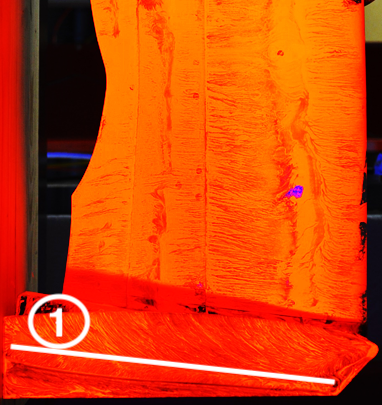}
\end{subfigure}
\hfill
\begin{subfigure}[b]{0.28\textwidth}
\includegraphics[width=1.0\textwidth]{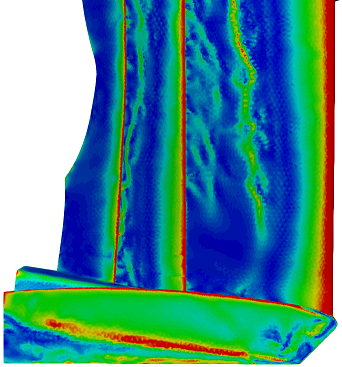}
\end{subfigure}
\hfill
\begin{subfigure}[b]{0.28\textwidth}
\includegraphics[width=1.0\textwidth]{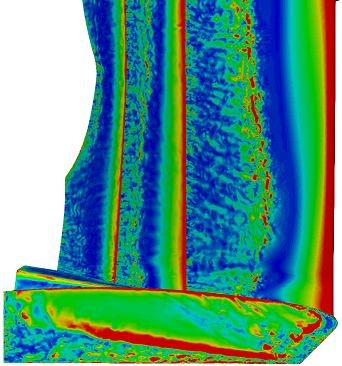}
\end{subfigure}
\hfill\null
\caption{Comparison of the flow visualization and wall shear stress for IFW (bottom view), with results from experiment (left), NM43 simulation (middle) and NM54 simulation (right)}
\label{fig: 8}
\end{figure}

\begin{figure}[bt]
\null\hfill
\begin{subfigure}[b]{0.265\textwidth}
\includegraphics[width=1.0\textwidth]{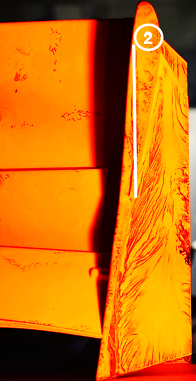}
\end{subfigure}
\hfill
\begin{subfigure}[b]{0.28\textwidth}
\includegraphics[width=1.0\textwidth]{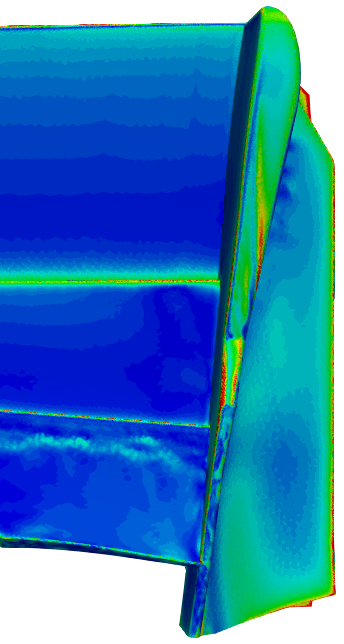}
\end{subfigure}
\hfill
\begin{subfigure}[b]{0.28\textwidth}
\includegraphics[width=1.0\textwidth]{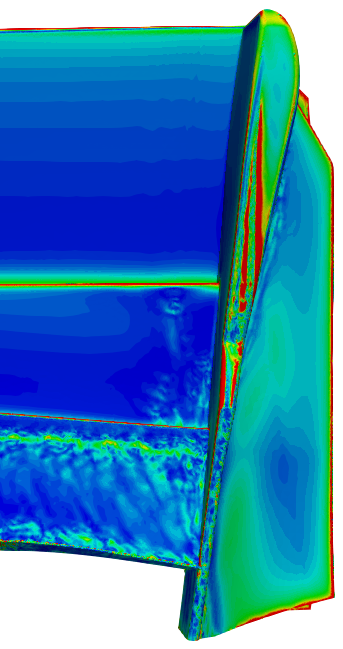}
\end{subfigure}
\hfill\null
\caption{Comparison of the flow visualization and wall shear stress for IFW (top view), with results from experiment (left), NM43 simulation (middle) and NM54 simulation (right)}
\label{fig:9}
\end{figure}

In Figures~\ref{fig: 8} and \ref{fig:9}, we compare the experimental flow  visualization with the wall shear stress magnitude from the Nektar++ simulations considering instantaneous plots at 10 $t_c$ for the NM43 and 11.75 $t_c$ for NM54. In Figure~\ref{fig: 8}, we first consider the bottom surface of the IFW with the simulations at NM43 and NM54 resolutions. All results show similar attachment lines on the footplate, indicating that the vortical structure positioning is likely to be consistent and the transition bubble in the main plane (first element of the cascade) is well captured. Increasing the polynomial order to NM54 leads to improvements in the prediction of the finer detailed features of the solution, better capturing the attachment line labelled 1 and also drawing out the more streaky structure of the attached flow on the main element. Although it is quite hard to identify the separation zones from the experimental data, careful consideration of  the images highlights a recirculation zone between (25-30\%) of the main element chord, considering the flow moving from right to left, and this recirculation zone is relatively well captured on the main element of the Nektar++ simulations.

Figure \ref{fig:9} also demonstrates a consistent prediction of the top edge and canard attachment lines for the two simulation cases. Comparing to NM43, the simulation at NM54 has a better resolved wall shear stress, indicating the separation lines on the vane (as indicated by the number 2) and flap, showing that increasing the polynomial order has a positive effect on the quality of the near wall flow prediction. 

\begin{figure}[bt]
\null\hfill
\begin{subfigure}[b]{0.325\textwidth}
\includegraphics[width=1.0\textwidth]{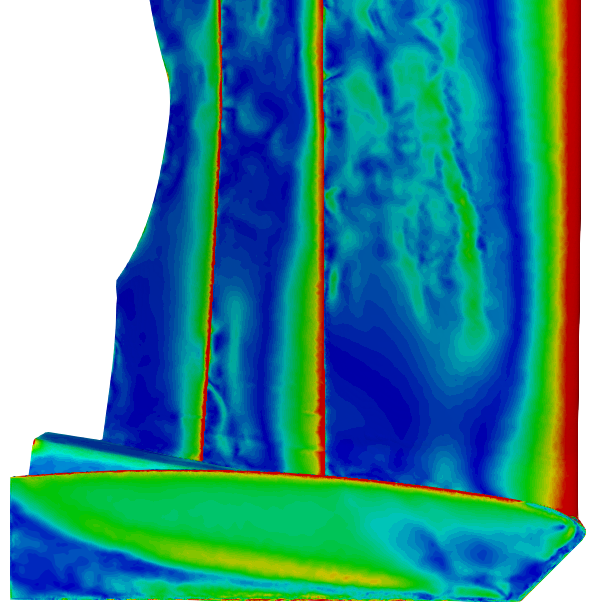}
\end{subfigure}
\hfill
\begin{subfigure}[b]{0.28\textwidth}
\includegraphics[width=1.0\textwidth]{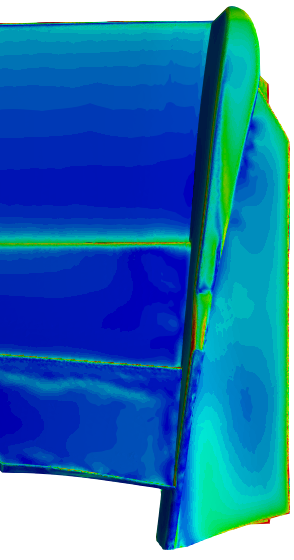}
\end{subfigure}
\hfill\null
\caption{Flow visualization results for IFW simulation considering \nth{3} polynomial expansion (NM32), with bottom (left) and top (right) views}
\label{fig:wssNM32}
\end{figure}

To confirm the discrepancy of NM32 resolution in predicting the flow structures, we present the flow visualization for this resolution on Figure~\ref{fig:wssNM32}, where most of the flow features found on experiments and higher-order simulations previously presented such as the vane separation line (indicated by 2) is only marginally captured.

\begin{figure}[bt]
\null\hfill
\begin{subfigure}[b]{0.325\textwidth}
\includegraphics[width=1.0\textwidth]{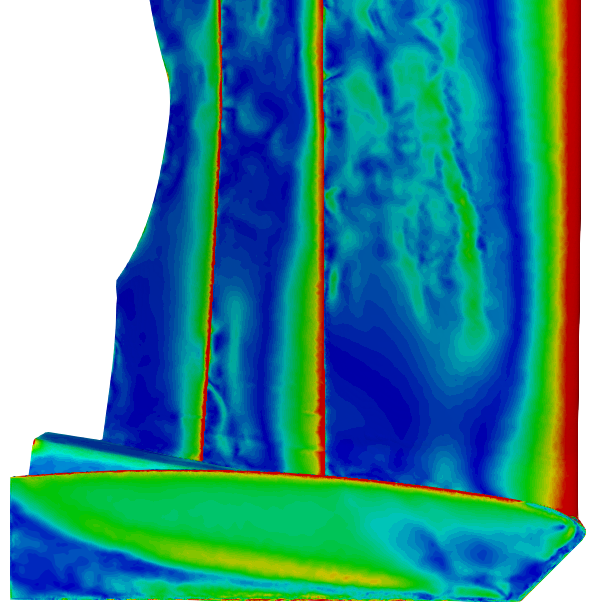}
\end{subfigure}
\hfill
\begin{subfigure}[b]{0.31\textwidth}
\includegraphics[width=1.0\textwidth]{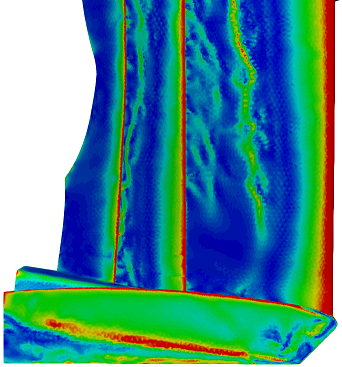}
\end{subfigure}
\hfill
\begin{subfigure}[b]{0.31\textwidth}
\includegraphics[width=1.0\textwidth]{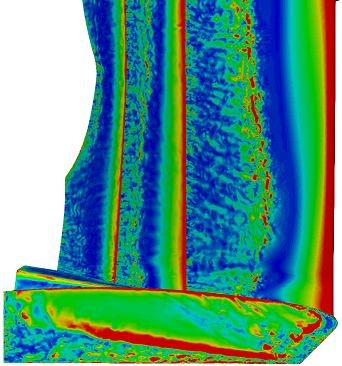}
\end{subfigure}
\hfill\null
\caption{Comparative flow visualization results for IFW simulation considering \nth{3} (NM32, left), \nth{4} (NM43, middle) and \nth{5} (NM54, right) polynomial expansion in order to highlight the major differences in resolution.}
\label{fig:wssbottomall}
\end{figure}

Comparing all three simulation resolutions in terms of the surface behaviour on each wing element presented on the bottom view in Figure~\ref{fig:wssbottomall}, for the NM54 case, flow separates from the main wing element then re-attaches all along the span. However for NM43, it does not seem to quite re-attach at the most outboard part of the main plane, where we observe a discontinued re-attachment line and increment of the separated flow region. At NM32 the flow is predominantly attached at about 2/3 of the span, similar to NM43 even with the separated flow region at the outboard part of the main wing element. At this point, NM54 presented similar flow topology as experimental results and NM43 and NM32, with similar results, slightly agreed. 

On the middle wing section, the flow on the NM54 case does re-attach, indicating similar profile of the main wing element. Once again the NM43 case manages to capture flow re-attachment but further downstream of the element, not so early and well-defined as on NM54 case. Finally, the NM32 is fully separated until the mid-span of the section, indicating that this resolution is not enough to capture the flow physics. Once again, NM54 results qualitatively well-captured the experimental results for this wing element, NM43 indicated similar flow physics with incorrect position and size and NM32 failed to represent experimental results.

Finally for the third wing element, where the gurney flaps are assembled, the flow only appears to be re-attaching in the NM54 case at the trailing edge of the wing element. On the footplate, NM54 and NM43 captures the vortex detached whereas NM32 only marginally captures this flow structure. We conclude that NM54 is the resolution required for this mesh setup to correctly capture the flow behaviour on the wing, although as we shall see in the following section, the NM43 can capture many feature of the time averaged flow.

\subsection{Comparison of PIV data}
\label{subsec:PIVPlaneComparativeResults}

\begin{figure}[bt]
	\begin{center}
	\includegraphics[width=0.35\textwidth]{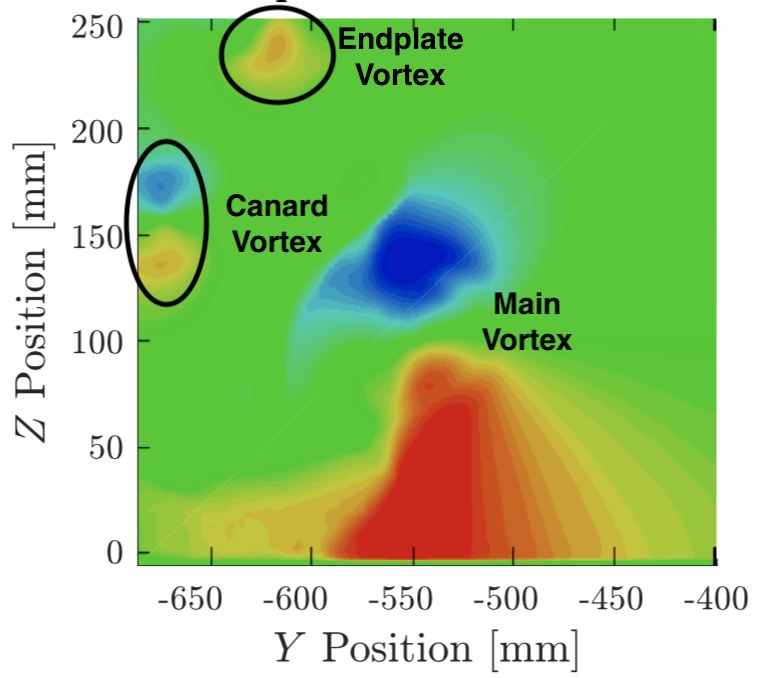}
	\caption{IFW vortices nomenclature definition. From top to bottom: endplate vortex (partially represented), canard vortex and main vortex.}
	\label{fig:vortex}
	\end{center}
\end{figure}

We now compare the numerical simulations against the PIV results on a few  planes at different downstream locations. Firstly, we identify the relevant vortex structures in the contour plots: the main, canard and endplate vortices are the dominant flow structures on the planes presented in this work and are depicted in Figure~\ref{fig:vortex}. 

\begin{figure}[bt]
\begin{subfigure}[b]{0.48\textwidth}
\begin{center}
\includegraphics[width=1\textwidth]{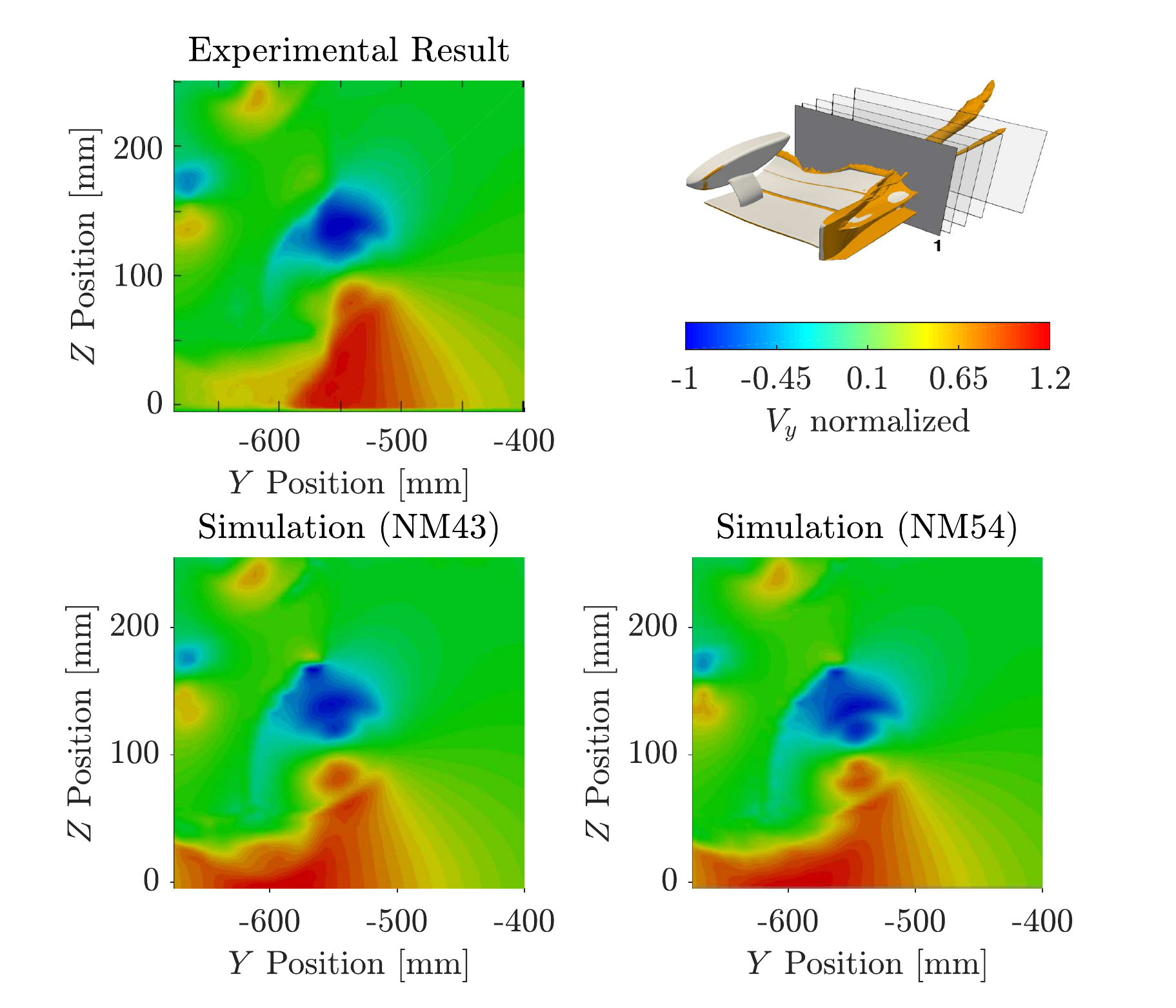}
\end{center}
\end{subfigure}
\begin{subfigure}[b]{0.48\textwidth}
\begin{center}
\includegraphics[width=1\textwidth]{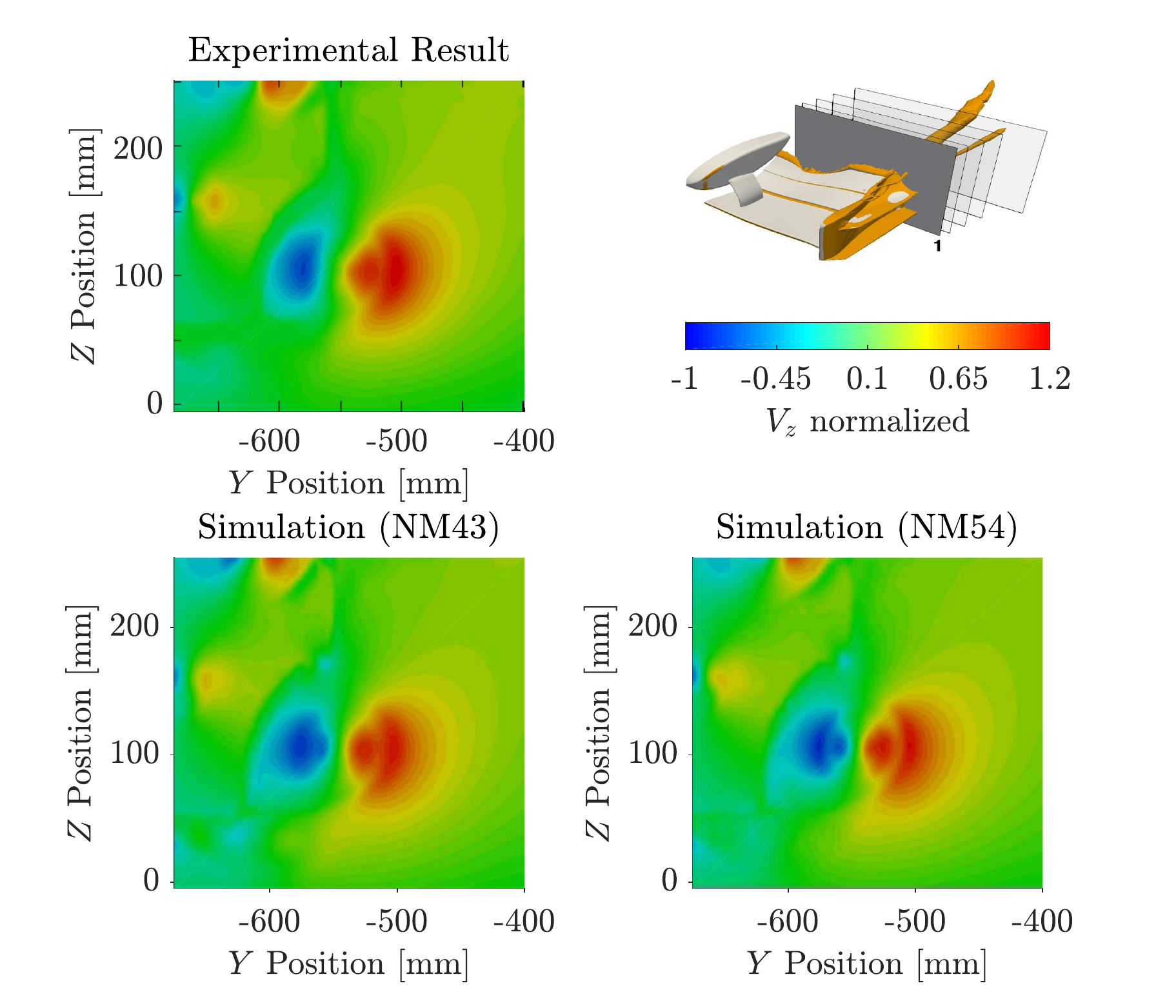}
\end{center}
\end{subfigure}
    \caption{Average normalized $V$ and $W$ velocity results from Nektar++ for \nth{4} (NM43) and \nth{5} (NM54) order polynomial expansion compared with experimental results for plane 1.}
\label{fig: n294}
\end{figure}

\begin{figure}[bt]
\begin{subfigure}[b]{0.48\textwidth}
\begin{center}
\includegraphics[width=1\textwidth]{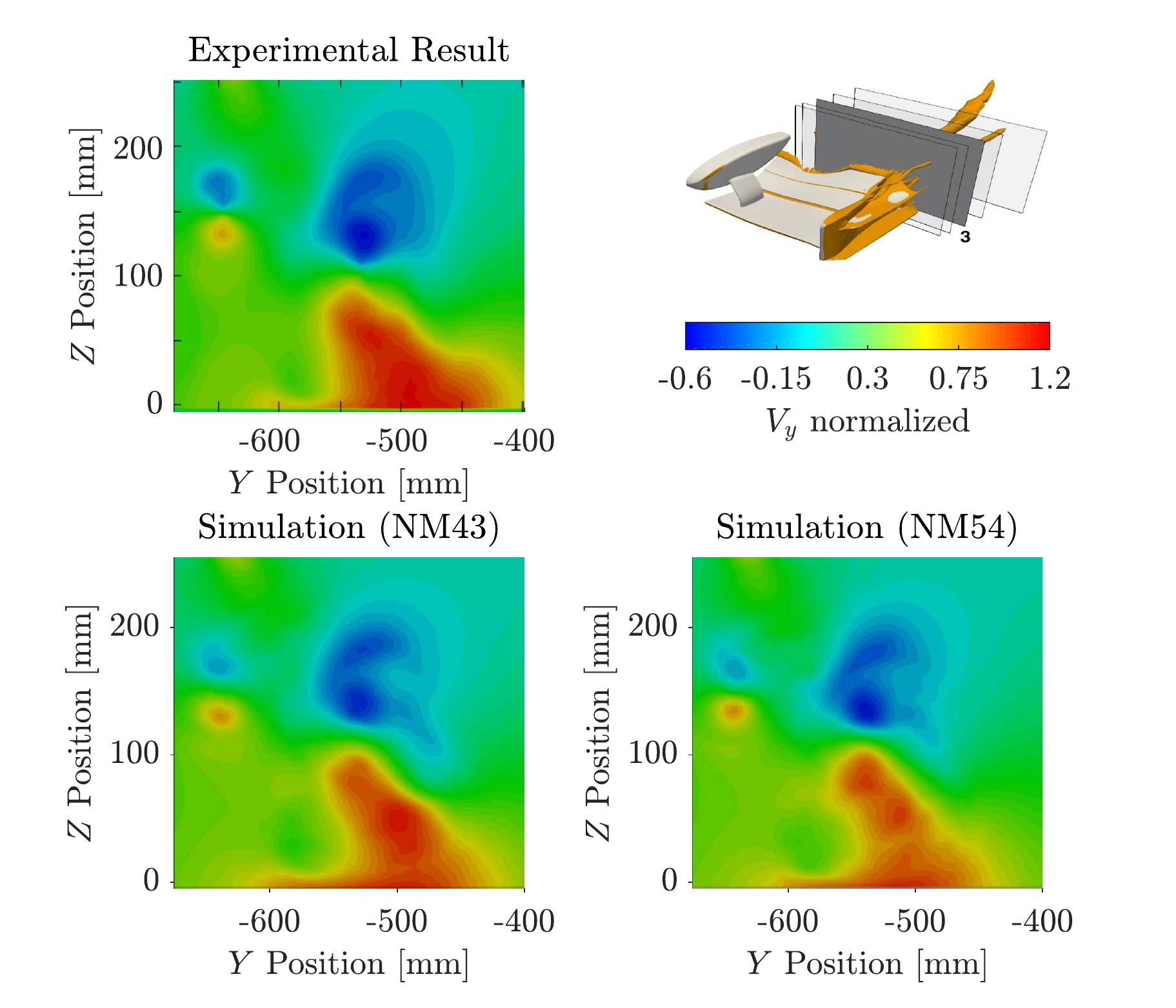}
\end{center}
\end{subfigure}
\begin{subfigure}[b]{0.48\textwidth}
\begin{center}
\includegraphics[width=1\textwidth]{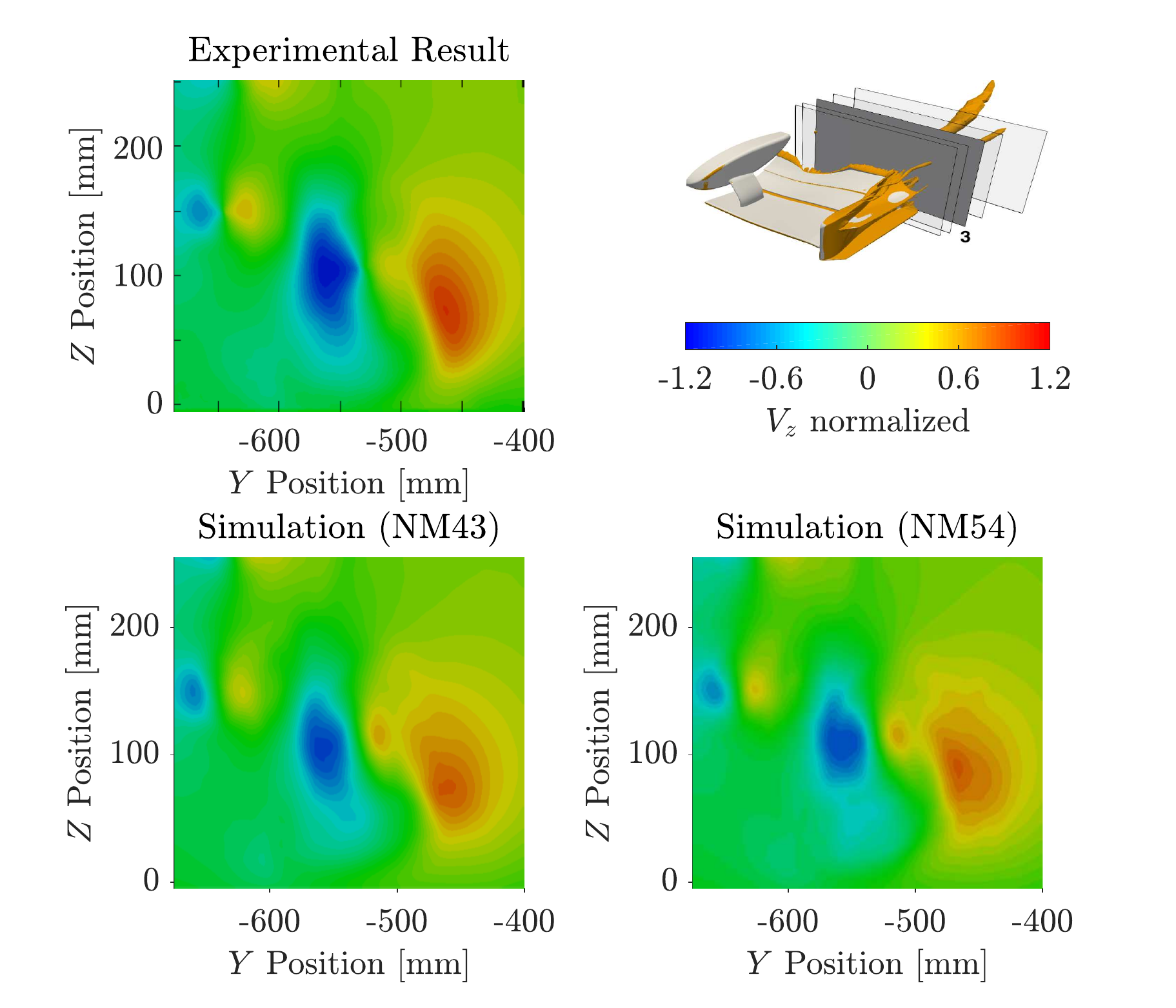}
\end{center}
\end{subfigure}
    \caption{Average normalized $V$ and $W$ velocity results from Nektar++ for \nth{4} (NM43) and \nth{5} (NM54) order polynomial expansion compared with experimental results for plane 3.}
\label{fig: n150}
\end{figure}

\begin{figure}[bt]
\begin{subfigure}[b]{0.48\textwidth}
\begin{center}
\includegraphics[width=1\textwidth]{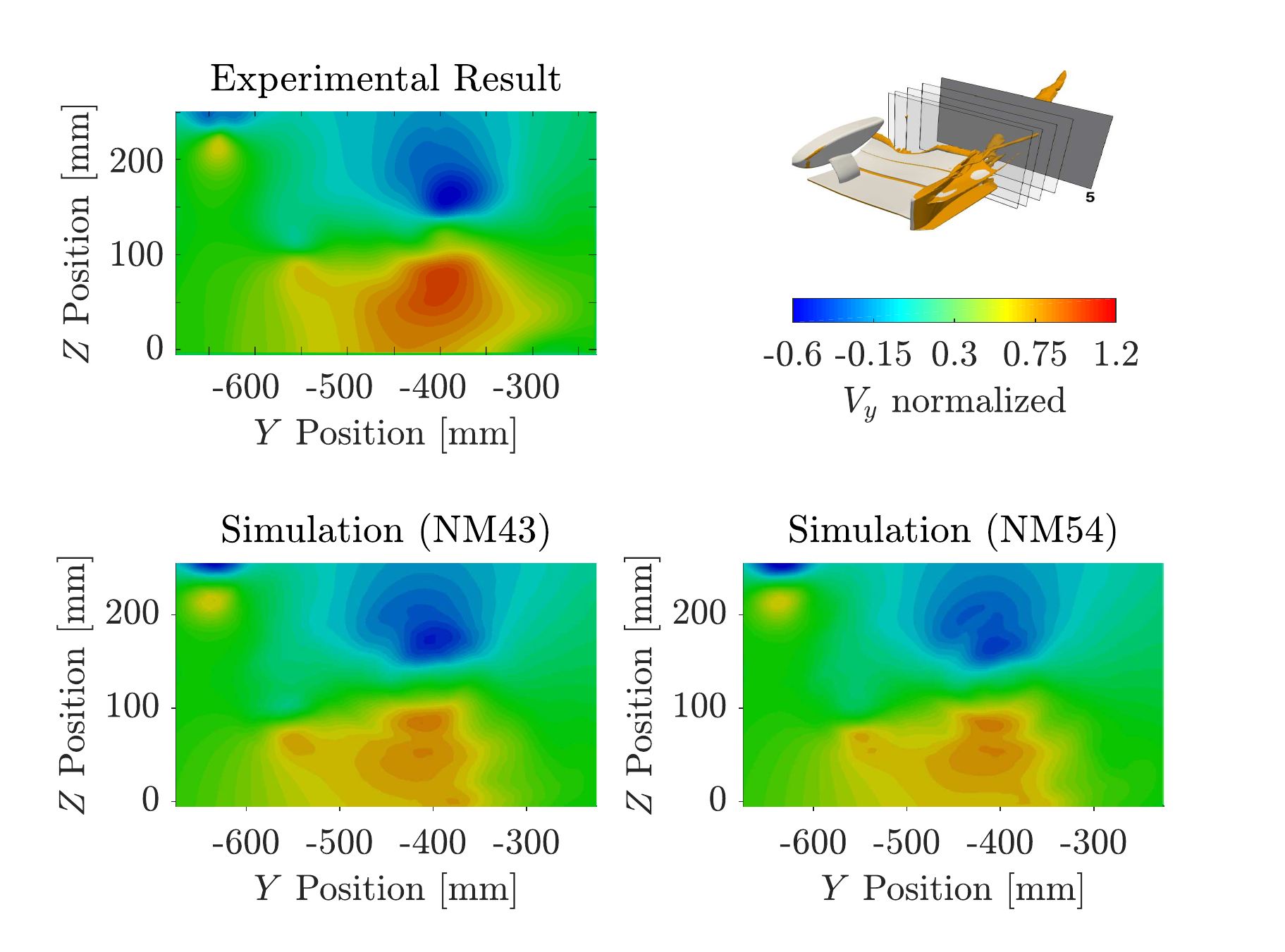}
\end{center}
\end{subfigure}
\begin{subfigure}[b]{0.48\textwidth}
\begin{center}
\includegraphics[width=1\textwidth]{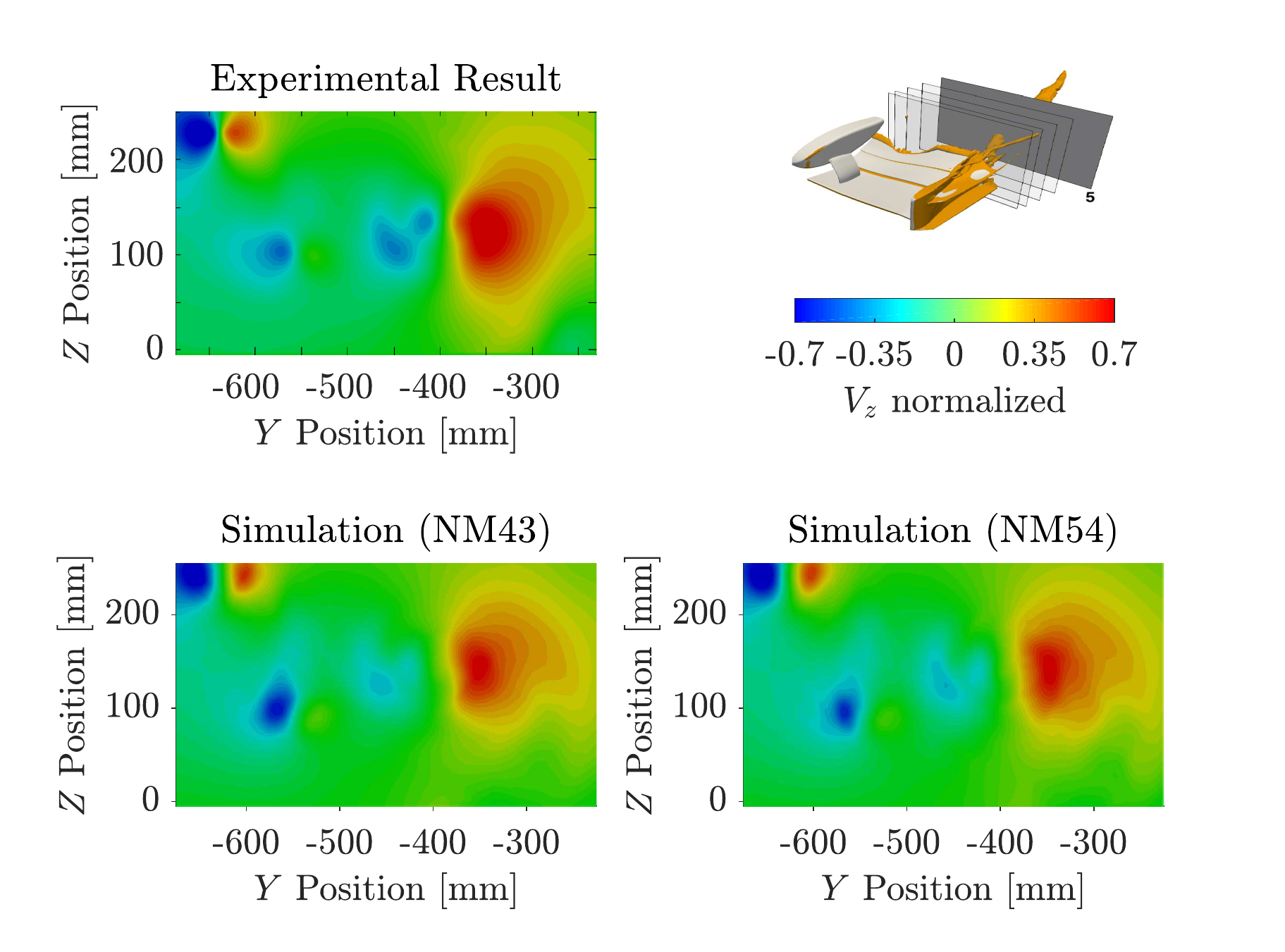}
\end{center}
\end{subfigure}
    \caption{Average normalized $V$ and $W$ velocity results from Nektar++ for \nth{4} (NM43) and \nth{5} (NM54) order polynomial expansion compared with experimental results for plane 5.}
\label{fig: p150}
\end{figure}

In the following, we present the comparisons for the plane closest to the wing (Figure~\ref{fig: n294}), the intermediate plane (Figure~\ref{fig: n150}), as well as a plane in the far field (Figure~\ref{fig: p150}), hence quantifying the prediction of the transport of vortices downstream. 

Both spanwise ($Y$) and vertical ($Z$) non-dimensionalised velocity components are presented. The location of each plane, as previously defined on \ref{subsubsec:PIVMeasurements}, is identified in the top right sub-plot. For each plane, we present time-averaged velocity results for the spanwise $V$ and vertical $W$ components, normalized by the freestream velocity $U$.

Considering the $V$ components shown in Figure~\ref{fig: n294} we observe a large blue (negative) region on top of a red triangular patch of $V$. These two features and their $W$ counterparts represent the main vortex arising from the inboard footplate of the endplate, denoting a region of counter clockwise rotating flow as we look at the plane from a downstream position: this is referred as the main vortex in the following. We also observe on the left side of Figure~\ref{fig: n294} two small circular regions, which are typically referred to as the canard vortex. The canard vortex is generated by the aerodynamic device mounted on the outboard side of the endplate (referred as canard), and is representative of the type of devices often introduced to provide additional load or shed vorticity. It has similar behaviour as the main vortex, with top circular region in blue (negative) and the lower portion is orange (positive), denoting similar counter clockwise rotation downstream for this vortex. The signature of this vortex is also clearly seen in $V$ and $W$ components shown on Figure~\ref{fig: n150}. 

The third relevant flow structure is observed on the top left side of Figure~\ref{fig: p150} and is defined here as the top-edge or endplate vortex. In the sampling window of this image we can only observe half the vortex as indicated by the orange (positive) circular region. The endplate vortex is generated by the pressure difference across the endplate according to \cite{pegrum2007experimental}.

We next consider the evolution of the three vortices, tracking their position as they travel downstream. The main vortex first appears with its core at around two-thirds of the spanwise location on plane 1 and the three elements on the wing contribute to its build up. As it travels downstream, it  gets diffused, as is evident from plane 3. Further downstream, by  plane 5, we observe that the main vortex core has shifted inwards (towards the symmetry plane) to around half of the span, which is consistent with the induced velocity from the image vortex in the plane of the floor. There is also evidence of the vortex moving upwards from the images of vertical velocity. 

Similarly, the  canard vortex  moves towards the symmetry plane, and its intensity is similar in  planes 1 to 3. Further downstream in plane 5, a very distinct movement in the negative $Z$ direction suggests that it is being ingested by the main vortex and this is perhaps most clear in the plots of the vertical velocity. For the endplate vortex, in spite of the absence of full results due to limitations in the PIV measurement window, we can still observe that it moves upwards in plane 3 and then downwards in plane 5, suggesting that the main vortex intensity influences the trajectory of the endplate vortex.

Figures~\ref{fig: n294}~ through ~\ref{fig: p150} show that both NM43 and NM54 polynomial orders have captured the main vortex, despite a slight under-prediction of its strength. In addition, the secondary features such as the endplate vortex and the canard vortex are also well captured by both numerical simulations. In short, for these resolutions both spectral/hp numerical simulations demonstrated good initial agreement with experimental results in terms of core position and intensity in the initial and downstream planes.

The difference between simulation results at \nth{4} and \nth{5} order can be observed in the sharper resolution of the smaller structures, which are less affected by diffusion at the higher \nth{5} order resolution. However, it is worth noting that the duration of averaging window used for NM54 is similar to NM43, at the same interval from 8.4 to 10 $t_c$  and we are confident that the planes where NM43 shows better agreement (namely plane 3) are mostly benefiting from a longer average and a more settled solution, as can also be seen from the lift traces. 


The results in the last plane, located more than one chord length away from the wing are then considered, where we recall the footplate vortex is merging into the main vortex. 


\subsection{Additional flow features}

\begin{figure}[bt]
\null\hfill
\begin{subfigure}[b]{0.44\textwidth}
\includegraphics[width=0.8\textwidth]{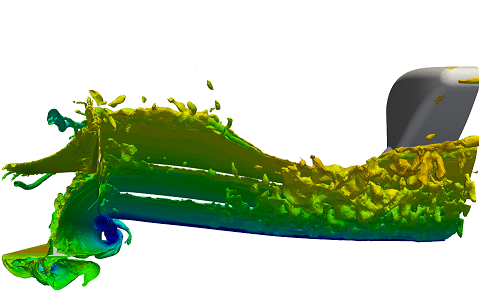}
\end{subfigure}
\hfill
\begin{subfigure}[b]{0.54\textwidth}
\includegraphics[width=0.8\textwidth]{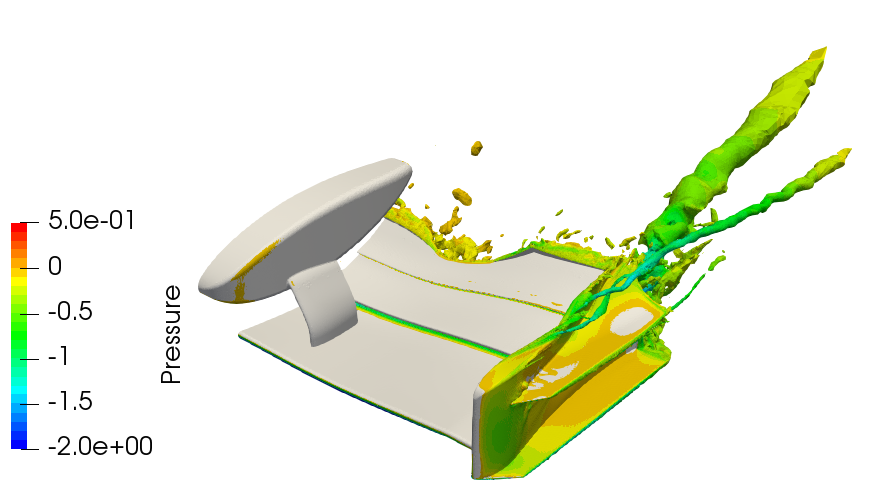}
\end{subfigure}
\hfill
\begin{subfigure}[b]{0.44\textwidth}
\includegraphics[width=0.8\textwidth]{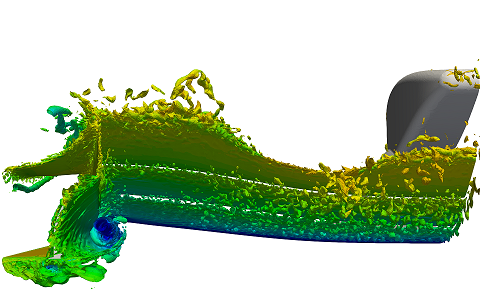}
\end{subfigure}
\hfill
\begin{subfigure}[b]{0.54\textwidth}
\includegraphics[width=0.8\textwidth]{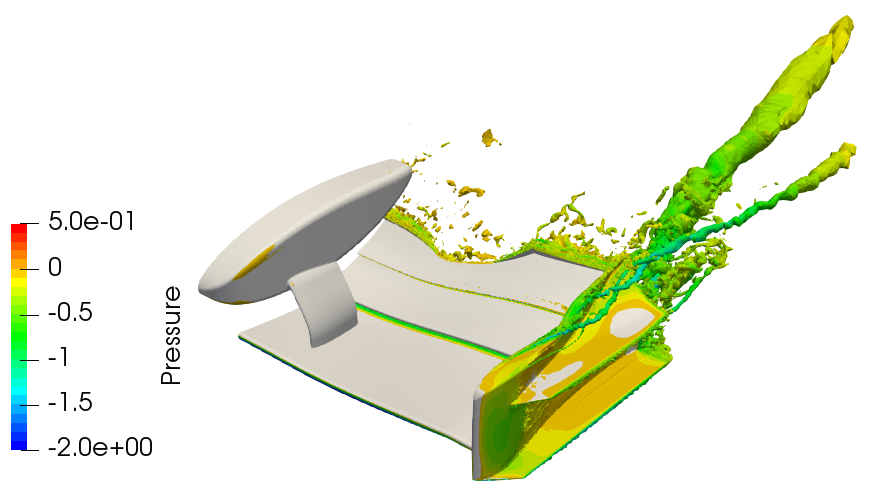}
\end{subfigure}
\hfill\null
\caption{Comparison of instantaneous iso-contours of $CP0 = 0$ coloured by pressure, with instantaneous plane 1 (left) and instantaneous full (right) views, for both NM43 (top) and NM54 (bottom) simulations}
\label{fig:IFW_isoCP0_inst}
\end{figure}

In the previous section we considered the time-averaged results, where the simulation with \nth{5} order polynomial expansion had only slightly better correlation with experimental results in terms of scales, intensity and the contour shape. These improvements are in fact the direct consequence of the better capturing a wider spectrum of flow structures, which can be highlighted when examining the instantaneous results instead. Figure~\ref{fig:IFW_isoCP0_inst} presents instantaneous iso-contour plots of $CP0=(p+0.5\rho u^2)/0.5\rho_\infty U_\infty^2$ comparing NM43 at 10 $t_c$ and NM54 at 11.75 $t_c$ in two scenarios: an upstream view from plane 1 location and a top-side view of the flow and scales over the wing. In the figure, comparing flow structures results of NM43 with that of NM54 simulation, we observe that NM54 does indeed capture smaller scales when compared NM43. It can also be seen that the partial flow separation visible close to the symmetry is significantly reduced when increasing the polynomial order, as was also demonstrated by the wall shear stress plots in Section~\ref{subsec:FlowVisualizationResults}.

\begin{figure}[bt]
\null\hfill
\begin{subfigure}[b]{0.4\textwidth}
\includegraphics[width=1.0\textwidth]{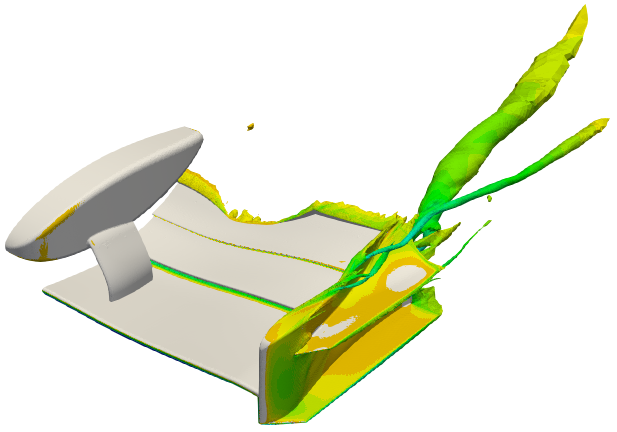}
\end{subfigure}
\hfill
\begin{subfigure}[b]{0.4\textwidth}
\includegraphics[width=1.0\textwidth]{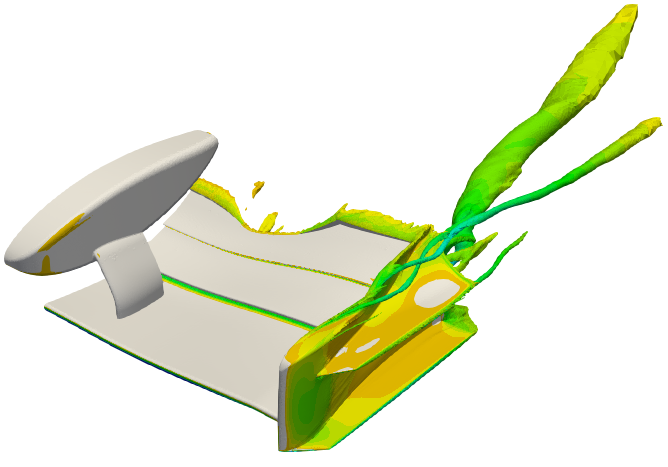}
\end{subfigure}
\hfill\null
\caption{Comparison of time averaged iso-contours of $CP0 = 0$ coloured by pressure, for NM43 (left) and NM54 (right) simulations.}
\label{fig:IFW_isoCP0_avg}
\end{figure}

Following on from the instantaneous images in Figure~\ref{fig:IFW_isoCP0_avg}, we show the time averaged iso-contour plots of $CP0=0$ over the IFW. A close inspection of the effect of the increment of resolution highlights that there is some change in the strength of the smallest vortices, such as the canard vortex and the initial formation of the endplate vortex, when comparing NM43 to NM54. Once again this was evidenced by the downforce increment, which was observed when enhancing the resolution from \nth{4} to \nth{5} order.

\section{Conclusions}
\label{sec:Conclusions}

This study presents a new test case for the automotive fluid dynamics community supported by experimental data, the Imperial Front Wing (IFW). The IFW geometry includes multiple wing elements under ground effect as well as other aerodynamic features such as gurney flaps, canard and foot plates, all of which are commonly used in present-day front wing designs. 

In terms of the locations of the vortex cores for the main, endplate and canard vortices, induced movements in both the spanwise and vertical directions were observed, as the vortices travel downstream and interact with each other. This type of complex geometry therefore leads to relatively intricate flow features that can be challenging to simulate and therefore obtain accurate prediction of the aerodynamic loads, transition lines locations and vortex intensities. We believe that the proposed geometry is therefore an interesting, aerodynamically challenging test case for CFD validation, especially for high-fidelity methods.

In this study we have compared the results of an uDNS/iLES simulation using third (NM32), fourth (NM43) and fifth (NM54) order spectral/hp element discretisation against experimental data. High-fidelity simulations were started from the NM32 resolution, and we first presented a comparison between results on the lift coefficient for NM32 and NM43, where results indicated discrepancy since at NM43 the downforce increased with time and NM32 followed the opposite trend (decreasing downforce). Lift coefficient results for the NM43 have converged around 8.4 $t_c$, whereas for NM54 it was still evolving, obtaining a stable region between the 11.5 and 11.75 $t_c$ and downforce increased by 4.5\% compared to NM43 results.

When complementing previous analysis with wall shear stress results, NM54 is the case that most closely correlates with the experimental results, capturing the re-attachment and flow features in all wing components. Results for NM43 indicate correct prediction of most of the features, except for the last wing component. The downforce increment on NM54 compared to NM43 is explained by the correct re-attachment of the flow on the wing elements, generating more downforce. For the NM32 case, flow features were marginally captured at the main wing element and the footplate and failed on capturing re-attachment on the other components, leading to the smallest downforce value and concluding that NM32 is an insufficient resolution for the proposed mesh.

The time-averaged results of NM43 and NM54 simulations are in relatively good agreement with the experimental PIV data. Consideration of the instantaneous solutions highlights that smaller scales are being captured in the higher resolution simulation, although these smaller scales are not having a significant effect on the time averaged evolution of the downstream vortices. 

We conclude that the minimum resolution for this mesh to correct capture the flow characteristics of vortical structures propagating downstream of the wing is NM43. The higher resolution does however help in capturing the less energetic vortices, improving the wall-bounded streaklines resolution and has therefore played a role in capturing a better downforce prediction.


\section*{Acknowledgments}

We acknowledge the work on the ExaFLOW Project for sponsoring this research and Imperial College London HPC, Archer and HP for the computational resources. We also acknowledge Dr. Oliver Buxton and Mr. Yushi Murai from the Department of Aeronautics - Imperial College London, for their work on the wind tunnel experiments. We are indebted to EPSRC through grant EP/L024888/1 for the National Wind Tunnel Facility (NWTF).

\subsection*{Author contributions}

Filipe F. Buscariolo - methodology development, experimental setup, data processing, correlation study, analysis and writing, reviewing and editing. 
Julien Hoessler - concept development, software implementation, methodology development, experimental setup, supervision, writing, reviewing and editing and sponsor.
David Moxey - software implementation, supervision, writing, reviewing and editing.
Ayad Jassim - computational resources, data analysis.
Kevin Gouder and Jeremy Basley - experimental setup and execution, data processing.
Gustavo R. S. Assi - supervision, writing reviewing and editing.
Spencer J. Sherwin - project coordinator, concept development, software implementation, writing, reviewing and editing and sponsor.

\subsection*{Financial disclosure}

None reported.

\subsection*{Conflict of interest}

The authors declare no potential conflict of interests.

\section*{Supporting information}

The following supporting information is available as part of the online article:

\noindent
\textbf{IFW CAD geometry and Nektar++ mesh files.}
{DOI: 10.14469/hpc/6049}

\FloatBarrier

\bibliography{wileyNJD-AMA}%

\end{document}